# Novel Local Radiomic Bayesian Classifiers for Non-Invasive Prediction of MGMT Methylation Status in Glioblastoma

Mihir Rao

Chatham High School
Chatham, NJ
November 2021

# Abstract

Glioblastoma, an aggressive brain cancer, is amongst the most lethal of all cancers. Expression of the O6-methylguanine-DNA-methyltransferase (MGMT) gene in glioblastoma tumor tissue is of clinical importance as it has a significant effect on the efficacy of Temozolomide, the primary chemotherapy treatment administered to glioblastoma patients. Currently, MGMT methylation is determined through an invasive brain biopsy and subsequent genetic analysis of the extracted tumor tissue. In this work, we present novel Bayesian classifiers that make probabilistic predictions of MGMT methylation status based on radiomic features extracted from FLAIR-sequence magnetic resonance imagery (MRIs). We implement local radiomic techniques to produce radiomic activation maps and analyze MRIs for the MGMT biomarker based on statistical features of raw voxel-intensities. We demonstrate the ability for simple Bayesian classifiers to provide a boost in predictive performance when modelling local radiomic data rather than global features. The presented techniques provide a non-invasive MRI-based approach to determining MGMT methylation status in glioblastoma patients.

# Table of Contents

# 1. Introduction

Glioma, particularly glioblastoma, and diffuse astrocytic glioma with molecular features of glioblastoma, are the most common and aggressive malignant primary tumor of the central nervous system in adults (World Health Organization Grade 4 astrocytoma) [1]. With extreme intrinsic heterogeneity in appearance, shape, and histology, glioblastoma patients have very poor prognosis, with the current standard care comprising of surgery, followed by radiotherapy and chemotherapy [2]. The efficacy of the first-line chemotherapy treatment, temozolomide, is in part dependent on the methylation status of the O6-methylguanine-DNA methyltransferase (MGMT) regulatory regions [3][4]. MGMT is a DNA repair enzyme, and the methylation of its promoter in newly diagnosed glioblastoma has been identified as a favorable prognostic factor and a predictor of chemotherapy response [2][4].

MGMT removes alkyl groups from compounds and is one of the few known proteins in the DNA Direct Reversal Repair pathway. Loss of the MGMT gene, or silencing of the gene through DNA methylation, may increase the carcinogenic risk after exposure to alkylating agents. High levels of MGMT in cancer cells can create resistance by blunting the therapeutic effect of alkylating agents and may be an important determinant of treatment failure. Thus, methylation of MGMT increases efficacy of alkylating agents such as temozolomide and improved median survival compared with patients with unmethylated gliomas (21.7 versus 12.7 months) [5][6][7]. Long-term follow-up has further substantiated the survival benefits [8][9][10].

The only reliable way to determine MGMT promoter methylation status requires analysis of glioma tissue obtained either via an invasive brain biopsy or following open surgical resection. MGMT status is inferred by histological analysis, as currently available non-invasive techniques are still too unreliable. Due to the absence of validation studies, there are major concerns related to evaluation of MGMT gene's promoter's methylation. Processing and assessment of data would require collaboration between molecular pathologists and neuro surgeons. The methods and cut-off definitions for determination of MGMT status also remain controversial [11]. Considerable attention has been recently dedicated to developing non-invasive, image-based diagnostic methods to determine MGMT promoter methylation status for glioblastoma.

Genetic changes may manifest as macroscopic morphological changes in the tumors that can be detected using magnetic resonance imaging (MRI), and can serve as non-invasive biomarkers for determining methylation of MGMT regulatory regions [12] [13] in the brain. Radiomic features have been used to extract textural tumor features from MRIs and have then been used to train and validate machine learning models [14 - 20]. Direct analysis of tumors has also been conducted using various deep-learning approaches [21 - 32]. Recently, methods have been outlined to conduct radiomic feature extraction at a local level using a sliding voxel-technique that calculates radiomic features for small fixed-size cubes throughout a three-dimensional MRI scan [33].

In this work, we utilize FLAIR-sequence MRIs, the PyRadiomics library [40], and local radiomic activation maps to develop simple Bayesian classifiers that provide the first results demonstrating the ability for local radiomics to boost classifier performance in comparison to global radiomic features. We also present new results outlining radiomic feature distributions with statistically significant differences across patient cohorts deemed by MGMT methylation status. The methods and results presented in this work provide novel radiomic insights into glioblastoma tumors and visual image features that can be non-invasively used to predict a glioblastoma patient's MGMT methylation status.

This paper is organized as follows. We first outline general radiomic feature extraction techniques, data acquisition, and the development of Bayesian classifiers in Sections 2.1, 2.2, and 2.3, respectively. We also describe Kernel Density Estimation (KDE), a facet of developing the Bayesian classifiers, in Section 2.4. We then present results pertaining to global radiomics in Sections 3-4, and local radiomics in Sections 5-6. Finally, we demonstrate local radiomics' ability to boost Bayesian classifier performance in Section 7 and conclusions in Section 8.

## 2. Methods

### 2.1 Radiomics in Oncology

#### 2.1.1 Global Textural Features

Radiomics is concerned with the extraction of quantitative metrics within medical images. It captures heterogeneity and shape from tissue and lesions and can be combined with other demographic, histologic or genomic data for clinical evaluations. Further, by capturing tumor radiomics properties over time additional insights can be gained into their changes, such as during patient treatment or surveillance. Radiomics also helps with the assessment of tissue heterogeneity and is of particular interest, as the degree of tumor heterogeneity is a prognostic determinant of survival [34][35][36]. While biopsies capture heterogeneity within only a small portion of a tumor and usually at just a single anatomic site, radiomics captures heterogeneity across the entire tumor volume. Studies have suggested that radiomic features are strongly correlated with heterogeneity indices at the cellular level [37][38] and therefore radiomic features are also associated with tumor aggressiveness [39].

Radiomic features can be roughly categorized as statistical (including histogram-based and texture-based), model-based, transform-based, and shape-based [40]. Here, we outline the radiomic feature types relevant to our work:

- **Histogram Features:** Statistical descriptors are based on the global gray level histogram and include gray-level mean, maximum, minimum, variance, percentiles, skewness, kurtosis entropy and energy.
- **Absolute Gradient**: Absolute gradient reflects the degree or abruptness of gray-level intensity fluctuations across an image and include gradient mean, variance, skewness, and kurtosis.
- **Gray-Level Co-occurrence Matrix (GLCM)**: GLCM is a second-order gray-level histogram and captures spatial relationships of pairs of pixels or voxels with predefined gray-level intensities, in different directions and with a predefined distance between the pixels or voxels.
- **Gray-Level Run-length Matrix (GLRLM).** The GLRLM provides information about the spatial distribution of runs of consecutive pixels with the same gray level, in one or more directions, in 2 or 3 dimensions.
- **Gray-Level Size Zone Matrix (GLSZM) and Gray-Level Distance Zone Matrix (GLDZM)**: GLSZM counts of the number of groups (so-called zones) of interconnected neighboring pixels or voxels with the same gray level form the basis for the matrix GLDZM not only assesses zones of interconnected neighboring pixels or voxels with the same gray level but requires them to be at the same distance from the region of interest (ROI) edge.

- **Neighborhood Gray-Tone Difference Matrix (NGTDM)**: NGTDM quantifies the sum of differences between the gray level of a pixel or voxel and the mean gray level of its neighboring pixels or voxels within a predefined distance.

- **Neighborhood Gray-Level Dependence Matrix (NGLDM):** NGLDM is based on the gray-level relationship between a central pixel or voxel and its neighborhood. Neighboring pixel or voxel within a predefined distance is regarded as being connected to the central pixel or voxel if it meets the dependence criterion in terms of a defined range of gray-level differences.

In this research, PyRadiomics, an open-source python package for the extraction of radiomic features from medical imaging, has been used for calculating global and local radiomics features [40]. Figure 1 shows a typical global radiomics workflow for feature extraction and modeling.

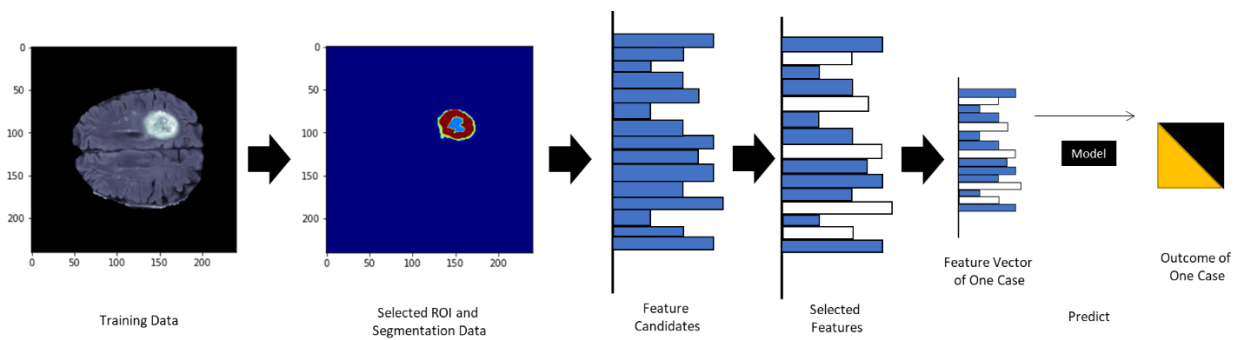

Figure 1. Global Radiomics Workflow for Feature Extraction and Modeling

### 2.1.2 Local Radiomics Techniques

Recently, Vuong et al. [33] presented the first work outlining methods for analysis of tumors using local radiomic activation maps. Local radiomics aims to extract the same features as those in global (default) radiomic feature extraction, but at specific voxel (three-dimensional/volumetric pixel) regions of the tumors. For each individual radiomic feature, the local radiomic value calculated within a specific voxel is then compared to the median of the global equivalent of that feature across all patients. If the local feature value is greater than or equal to the global median for that feature, that voxel is considered "activated" for that feature type in that specific patient's MRI. Contrarily, if the local value is less than the global median, it is considered "non-activated". Figure 2 illustrates the process of progressing from global radiomics, to local radiomic feature value maps, and finally to local radiomic activation maps.

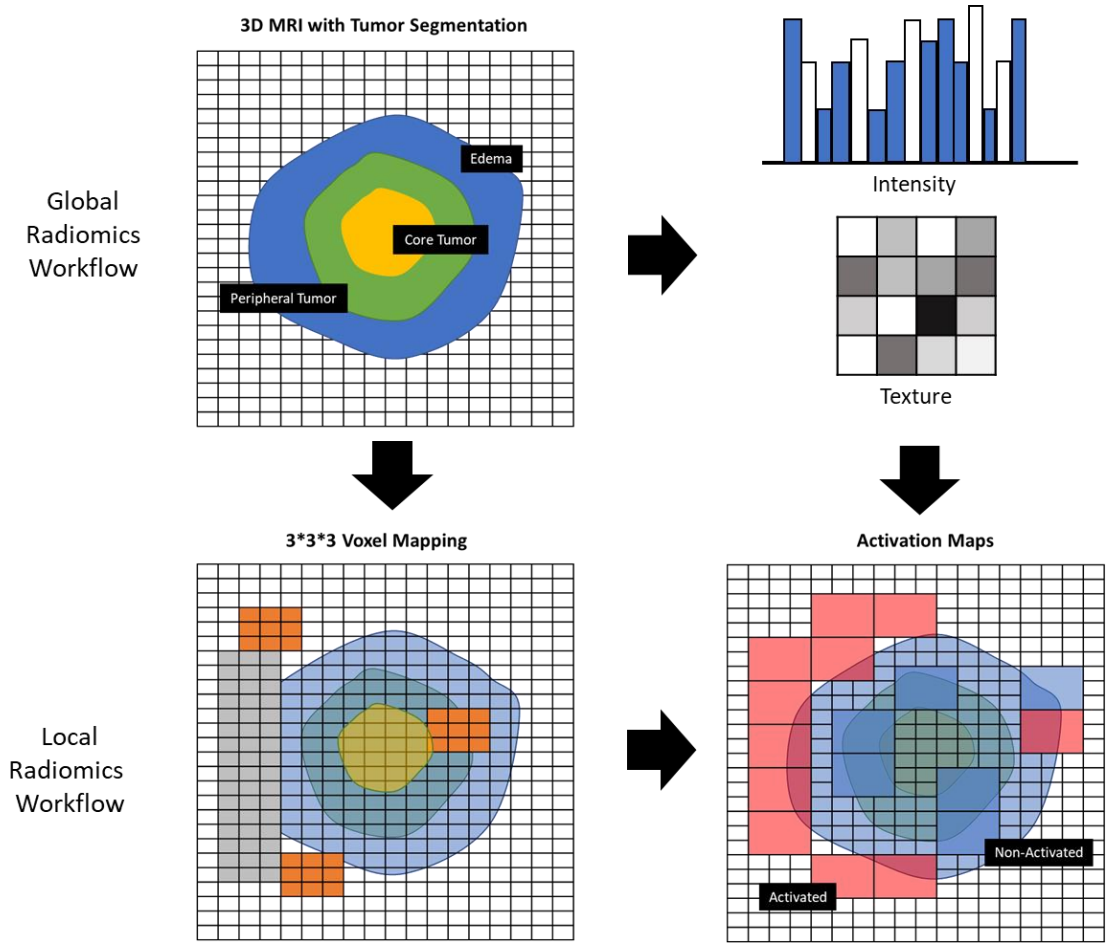

**Figure 2. Global to Local Radiomics Workflow for Feature Value and Activation Maps**

In this work, we implement a three-dimensional cropping-technique by forming a minimum-size cube to isolate the general tumor region of the full brain MRI while minimizing the amount of non-tumor brain tissue present.

Once the cube is formed, we resize it to a uniform size of 33x33x33 voxels (35,937 total voxels) that allows for fast calculations while maintaining MRI details and nuances in brain tissue architecture and appearance. In this cropped region of the MRI, we implement a sliding-cube technique that slides a 3x3x3 voxel cube (27 total voxels in the sliding-cube) across the entire tumor-containing cube and calculate local radiomic values for those cubes that have at least 1 tumor tissue voxel out of the 27 voxels of brain tissue based on the segmentation data of each patient. The calculated local radiomic values within each of the 3x3x3 voxel cubes are then considered with the global median radiomic value for the feature in question in order to determine each 3x3x3 cube's "activated" or "non-activated" status.

**2.2 Data**

The dataset used in this work is from the BraTS 2021 challenge [41-46] and has been collected from institutions around the world as part of a decade-long project to advance the use of AI in brain tumor diagnosis and treatment. The BraTS 2021 dataset utilizes multi-institutional pre-operative baseline multi-parametric magnetic resonance imaging (mpMRI) scans, and focuses on the evaluation of state-of-the-art methods for segmentation of intrinsically heterogeneous brain glioblastoma sub-regions in mpMRI scans (Task 1 of the challenge) and classification methods to predict the MGMT promoter methylation status (Task 2 of the challenge). Figure 3 shows FLAIR 3D MRI scans with segmentation data for sample patient.

For the purposes of this work, we take advantage of the fact that a vast majority of the same patients were present in the data for each individual task, allowing us to collect segmented FLAIR-sequence. MRI tumor data and the associated MGMT methylation status for patients that were present across the two challenge tasks based on their unique patient identification numbers. By doing so, we compiled a comprehensive dataset containing 577 unique patients, 301 of whom had methylated MGMT (MGMT 1) and 276 of whom had unmethylated MGMT (MGMT 0). In this work, we also consider the whole tumor from the provided segmentation data, which consists of three tumor regions: the tumor core, the peripheral tumor, and edema surrounding the tumor.

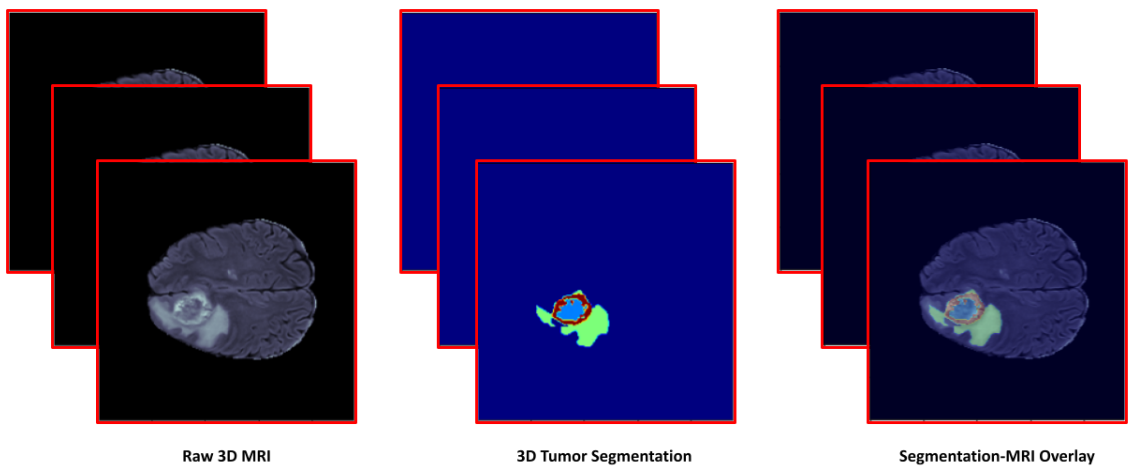

**Figure 3. FLAIR 3D MRI scans with Segmentation Data for Sample Patient**

## 2.3 Bayesian Classification

Bayesian Classification [47] are a set of probabilistic classifiers that aim to process, analyze, and categorize new data based on a finite set of known data. Assume a two-class dataset containing $N(A)$ individuals belonging to class A and $N(B)$ individuals belonging to class B. The relative frequency of class A in the dataset is given by the number of individuals belonging to class A divided by the total number of individuals in the dataset:

$$P(A) = \frac{N(A)}{N(A) + N(B)} \tag{1}$$

Likewise, we can calculate the relative frequency of class B in the dataset:

$$P(B) = \frac{N(B)}{N(A) + N(B)} \tag{2}$$

Assuming that the belonging of an individual to one class is a mutually exclusive to the individual being described by some feature $x$, Bayes' Theorem provides the probability of an individual belonging to class A for some given descriptive feature $x$:

$$P(A|x) = \frac{P(x|A) * P(A)}{P(x)} \tag{3}$$

This also extends to the probability of an individual belonging to class B:

$$P(B|x) = \frac{P(x|B) * P(B)}{P(x)} \tag{4}$$

In its simplest form, a classification model aims to receive an input and produce a prediction of a class as an output based on the input. If we consider our descriptive feature, $x$, as an input, then a classifier based on Bayes' Theorem can be constructed by checking for the satisfaction of an inequality that would result in the classifier predicting one class over the other in a two-class dataset. In our dataset of classes, A and B, a classifier would predict class A for some input value $x$ if the probability of an individual belonging to class A is higher than that individual belonging to class B given $x$:

$$P(A|x) \geq P(B|x) \tag{5}$$

With Bayes' Theorem, we can substitute in for this inequality and write it in its expanded form:

$$\frac{P(x|A) * P(A)}{P(x)} \geq \frac{P(x|B) * P(B)}{P(x)} \tag{6}$$

$$P(x|A) * P(A) \geq P(x|B) * P(B) \tag{7}$$

In equation 6, we can immediately observe that division by $P(x)$ is present on both sides, suggesting that we actually do not need to know $P(x)$ to make a classification with our Bayesian classifier. The inequality can, thus, be written in its final simplified form without the denominators (equation 7). In the context of our work, this Bayesian classification model is used to take in an input $x$ (a quantitative value pertaining to extracted radiomic features from a patient) and then predict if that patient belongs to class A (MGMT 1) or class B (MGMT 0). The $P(x|A)$ and $P(x|A)$ components of the Bayesian classification inequality, however, require continuous probability distributions, and the finite nature of our patient dataset means that we only have discrete probabilities for bins, or ranges, of data. To estimate the continuous probability distribution that best represents our discrete histogram-form data, we utilize Kernel Density Estimation, which is outlined in the subsequent sub-section.

**2.4 Kernel Density Estimation**

Kernel density estimation (KDE) is a non-parametric way to estimate the probability density function of a random variable $X$. It is a statistical data smoothing method in which inferences about the finite and discrete data are made, and is defined as the sum of a kernel function on every distinct data point in a distribution. While a histogram counts the number of data points in somewhat arbitrary regions, a KDE works by plotting out the data and creating a continuous curve of the distribution composed of discrete data. The curve is calculated by weighing the distance of all the points in each unique location along the distribution's horizontal axis.

A critical determinant of a KDE's ability to well-represent the structure of a finite distribution is its bandwidth parameter. The bandwidth of a kernel function plays an important role in fitting the data as it determines the "smoothness" of the KDE-estimated curve. The choice of a bandwidth for a kernel density estimator is important in finding a suitable density estimate. Too narrow a bandwidth can lead to a high-variance estimate and too wide a bandwidth can lead to a high-bias estimate where the structure in the data is washed out by the wide KDE kernel. In this work, we utilize Sci-Kit Learn's GridSearchCV [49] class to optimize a Gaussian Kernel Density Estimator [48] for the bandwidth parameter before further utilizing it when developing and making predictions with our Bayesian classifiers.

## 3. Global Radiomic Feature Extraction

We begin our collection of results by first extracting 93 different PyRadiomics textural radiomic features from our cohort of 577 glioblastoma patients. These 93 features types include both first and second order textural features. Once global radiomic feature extraction was complete for each patient's tumor, each patient had an associated 93-length list of global radiomic values. As a normalization step, the raw radiomic value ($x$) for a specific feature and specific patient was converted to a Z-score using the mean radiomic value for that specific feature across all patients $\mu$ and the standard deviation of the distribution of that specific feature across all patients $\sigma$:

$$Z = \frac{x - \mu}{\sigma}$$

To select features that would be used to develop Bayesian classifiers, we conducted t-tests for each of the 93, now Z-normalized, features in which we looked for a statistically significant difference ($p < 0.05$) between that specific feature's distributions for MGMT 1 and MGMT 0 patients, respectively. Table 1 enumerates the names of the 46 global radiomic features that yielded statistically significant differences between MGMT 1 and MGMT 0. Figure 4 depicts a sample of 9 distribution pairings out of these 46 features.

## 4. Bayesian Classification of Z-Normalized Global Features

Using the simple Bayesian classifier from equation 7, we can classify patients as either MGMT 1 or MGMT 0 using a patient's Z-normalized value for a specific global radiomic feature as the model input x. In this way, we create 46 distinct Bayesian classifiers for each of the 46 global radiomic features that demonstrated statistically significant differences between MGMT 1 and MGMT 0 patients. We utilize a 10-fold stratified cross-validation approach to split the dataset into 10 equal sized folds with close to perfectly consistent class balance across folds. During the 10 folds of modelling and validation of each feature-specific Bayesian classifier, several steps occur in sequence:

| Feature | p-value | Feature | p-value |
|---|---|---|---|
| glcm_Correlation | 0.033872 | glszm_GrayLevelNonUniformityNormalized | 0.044348 |
| glcm_Imc2 | 0.018582 | glszm_ZoneEntropy | 0.004969 |
| glcm_DifferenceVariance | 0.017768 | glszm_SmallAreaEmphasis | 0.008056 |
| glcm_Idm | 0.014497 | glszm_SmallAreaLowGrayLevelEmphasis | 0.006291 |
| glcm_Contrast | 0.013887 | glrlm_ShortRunHighGrayLevelEmphasis | 0.008866 |
| glcm_Id | 0.012793 | glszm_SmallAreaHighGrayLevelEmphasis | 0.007020 |
| firstorder_RobustMeanAbsoluteDeviation | 0.012309 | glrlm_HighGrayLevelRunEmphasis | 0.008385 |
| firstorder_InterquartileRange | 0.009808 | glcm_SumEntropy | 0.001461 |
| glcm_Autocorrelation | 0.009629 | glrlm_RunEntropy | 0.000681 |
| firstorder_MeanAbsoluteDeviation | 0.006361 | gldm_DependenceNonUniformityNormalized | 0.037342 |
| firstorder_Minimum | 0.003252 | gldm_SmallDependenceEmphasis | 0.014324 |
| glcm_DifferenceAverage | 0.002711 | glszm_HighGrayLevelZoneEmphasis | 0.006252 |
| glcm_DifferenceEntropy | 0.002379 | glszm_SizeZoneNonUniformityNormalized | 0.009595 |
| glcm_JointAverage | 0.001344 | ngtdm_Complexity | 0.025920 |
| firstorder_Entropy | 0.001301 | glcm_SumAverage | 0.001344 |
| firstorder_Range | 0.001102 | gldm_SmallDependenceHighGrayLevelEmphasis | 0.015046 |
| glcm_InverseVariance | 0.000484 | glszm_ZonePercentage | 0.014414 |
| firstorder_10Percentile | 0.000072 | glcm_JointEntropy | 0.003470 |
| firstorder_Maximum | 0.000069 | gldm_DependenceEntropy | 0.030074 |
| firstorder_Median | 0.000065 | gldm_LargeDependenceHighGrayLevelEmphasis | 0.026357 |
| firstorder_Mean | 0.000056 | gldm_HighGrayLevelEmphasis | 0.008651 |
| firstorder_90Percentile | 0.000055 | glcm_MCC | 0.018294 |
| firstorder_RootMeanSquared | 0.000052 | glrlm_LongRunHighGrayLevelEmphasis | 0.006542 |

**Table 1. Statistically significant Global Radiomics Features across MGMT 1 and MGMT 0 for (p < 0.05)**

- Two Kernel Density Estimators (KDEs) are optimized for the bandwidth parameter given the raw input feature distribution using a brute force grid search within a manually set domain of bandwidth values. The two KDEs are for the MGMT 1 and MGMT 0 distributions of a specific global radiomic feature, respectively.

- $P(x|A) * P(A)$ and $P(x|B) * P(B)$ from equation 7 are calculated using the two bandwidth-optimized KDEs across a domain of hypothetical input Z-normalized feature values that has a high level of precision (for these global radiomic Bayesian classifiers, we utilize 3 decimal places of the Z-normalized global radiomic feature value).

- The Bayesian classifier's prediction of MGMT 1 or MGMT 0 is determined for all feature values within the utilized domain of hypothetical input Z-normalized feature values using the calculated values for $P(x|A) * P(A)$ and $P(x|B) * P(B)$ at each input value $x$ in the hypothetical domain and the inequality logic of equation 7.

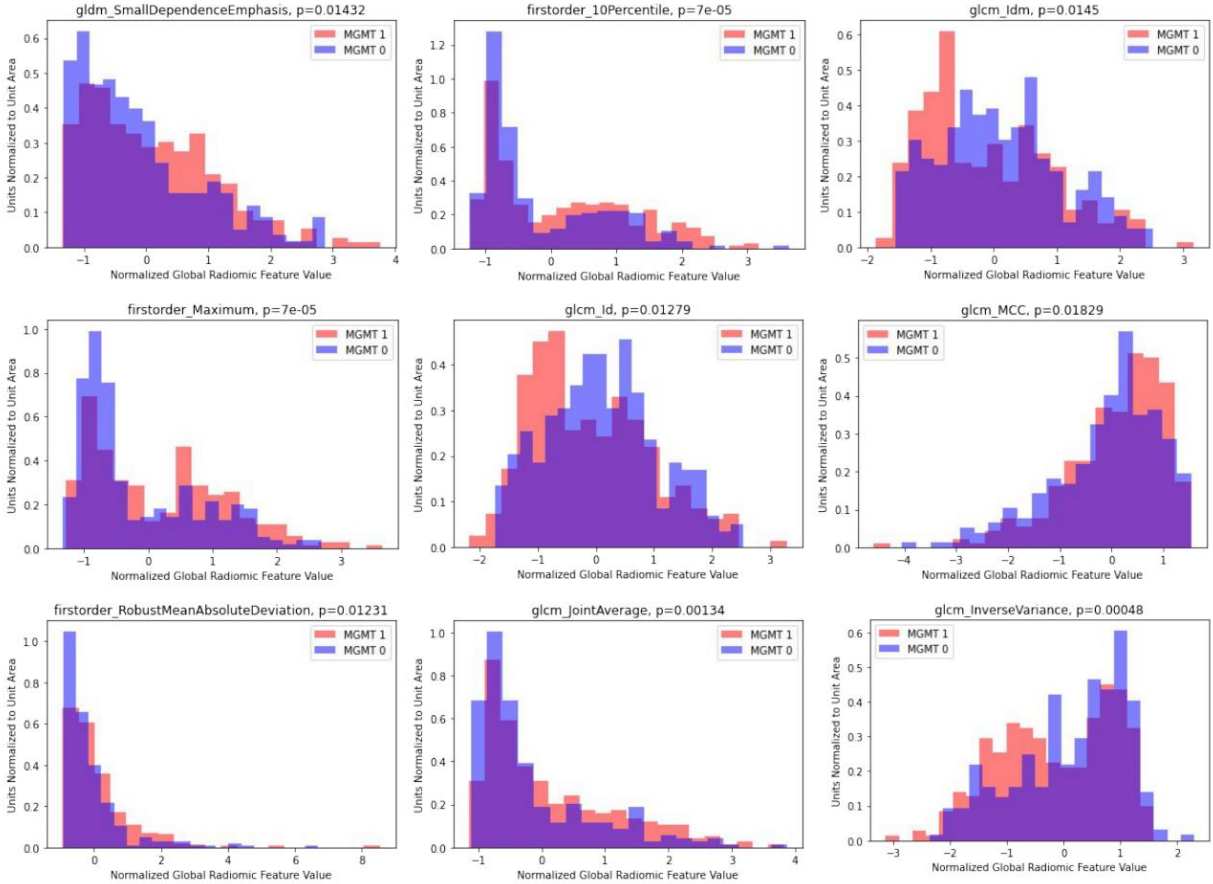

**Figure 4.  Distribution for MGMT 1 and MGMT 0 pairing for Sample Global Radiomics Features**

These three steps in the development of a Bayesian classifier for a global are illustrated in Figure 5, which shows visual representations of each of the three steps for two examples of global radiomic features.

Note that the Bayesian classifiers reflected in Figure 5 have been modelled across all 577 patients, but the 10-fold stratified cross-validation process would model a unique 90% of the dataset in each unique fold while using the remaining 10% of the patients as a validation set within the fold. The results of the cross-validation process for the Bayesian classifiers of each of the 46 features are provided in Section 7 of this paper.

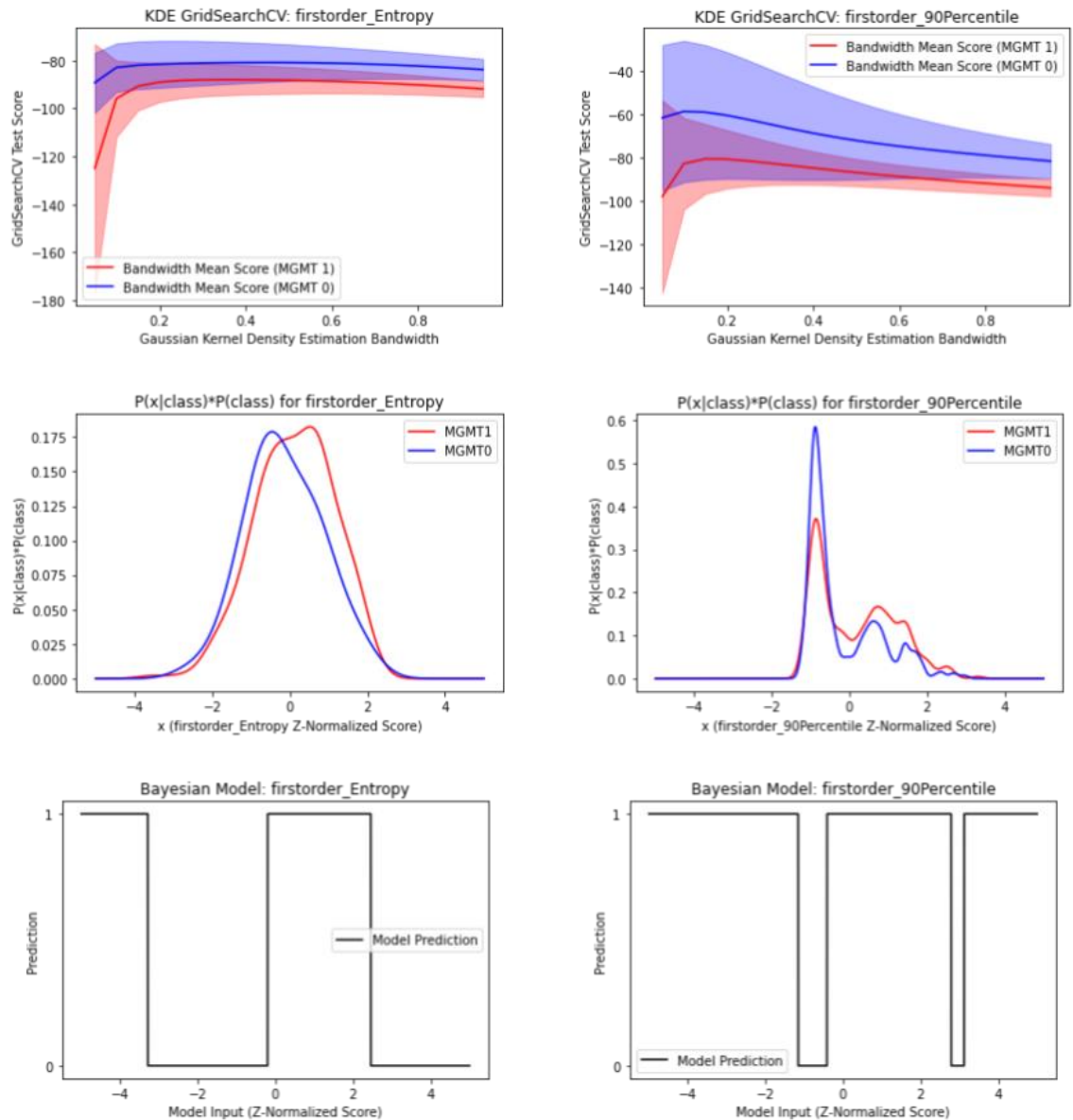

**Figure 5. Bandwidth Optimization and Bayesian Classifier for Global Radiomics Based Predictions**

## 5. Local Radiomics Activation Maps

Following the local radiomic workflow described in Section 2.1.2, we created local radiomic feature value maps and subsequent activation maps for all 577 patients across all 93 PyRadiomics textural features that were used in this work. Figure 6 offers an example of such a local radiomic value map and activation map pairing for the PyRadiomics first order-Kurtosis feature in a sample patient.

For each patient, the completed feature-specific local activation map was used to calculate the percentage of "activated" voxels in their tumor %$A$:

$$\%A = \frac{A}{A + N}$$

where A is the number of activated voxels and N is the number of non-activated voxels for that feature. These percentages serve as a normalized form of local radiomic data, as opposed to the raw radiomic value taken at a 3x3x3 voxel-level.

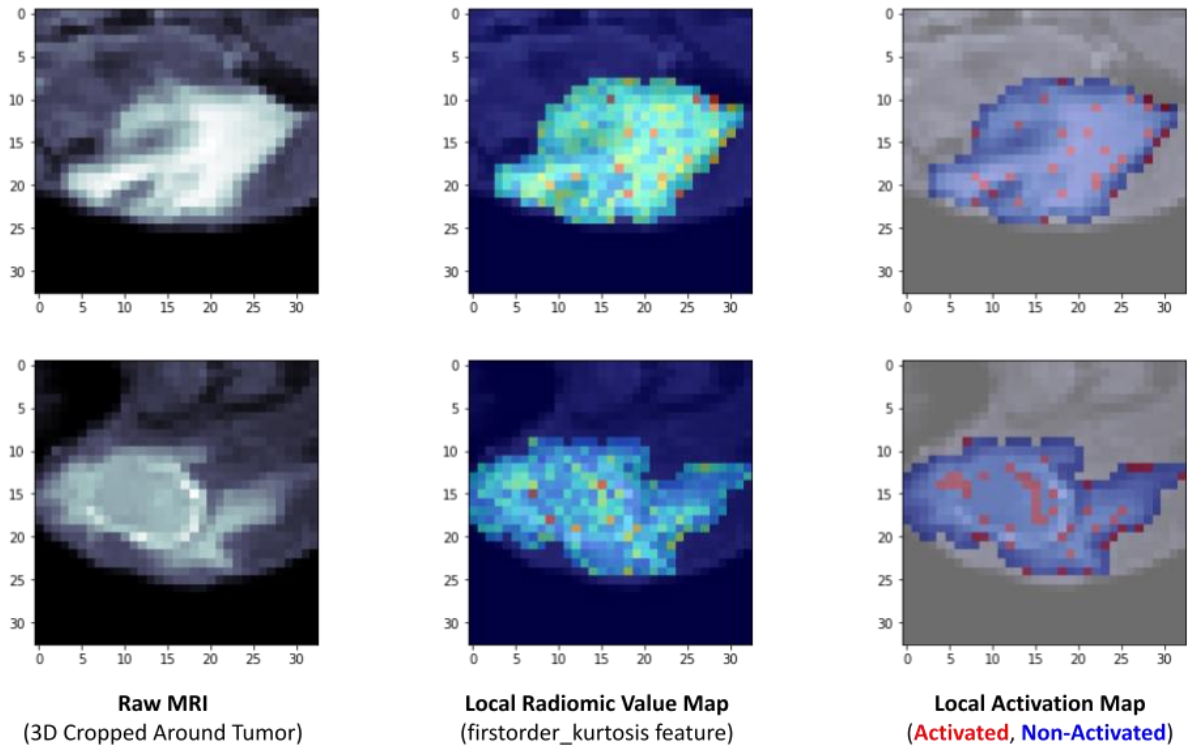

**Figure 6. Local Radiomics Value Map and Activation Maps for First Order Kurtosis Radiomics Feature**

Much like the t-test-based feature selection conducted on global radiomic data, we identify radiomic features that have statistically significant differences between their MGMT 1 and MGMT 0 patient distributions of the local radiomics-based percentage activation calculation. Table 2 lists the 62 out of 93 features that were found to be statistically significant through the local radiomic approach. Figure 7 shows a sample of 9 distribution pairings out of these 62 significant features.

| Feature | p-value | Feature | p-value |
|---|---|---|---|
| glcm_DifferenceEntropy | 0.016125 | glszm_HighGrayLevelZoneEmphasis | 0.003766 |
| firstorder_Skewness | 0.009414 | glrlm_LongRunEmphasis | 0.000175 |
| firstorder_Minimum | 0.008749 | gldm_SmallDependenceHighGrayLevelEmphasis | 0.000474 |
| firstorder_Range | 0.007492 | glrlm_GrayLevelNonUniformityNormalized | 0.001874 |
| glcm_Imc1 | 0.004829 | glszm_SmallAreaHighGrayLevelEmphasis | 0.002945 |
| firstorder_Kurtosis | 0.003394 | gldm_GrayLevelVariance | 0.000505 |
| glcm_JointAverage | 0.003363 | ngtdm_Complexity | 0.003547 |
| glcm_SumAverage | 0.003363 | ngtdm_Strength | 0.000811 |
| glcm_Autocorrelation | 0.003271 | glrlm_RunPercentage | 0.000272 |
| firstorder_Entropy | 0.002675 | gldm_DependenceNonUniformityNormalized | 0.000918 |
| glcm_DifferenceVariance | 0.002289 | glszm_LargeAreaLowGrayLevelEmphasis | 0.000080 |
| firstorder_Uniformity | 0.001864 | glrlm_LongRunHighGrayLevelEmphasis | 0.006341 |
| glcm_ClusterProminence | 0.001457 | glrlm_RunVariance | 0.000905 |
| glcm_ClusterTendency | 0.000885 | glrlm_ShortRunHighGrayLevelEmphasis | 0.002588 |
| firstorder_Variance | 0.000524 | glrlm_RunLengthNonUniformityNormalized | 0.000286 |
| firstorder_MeanAbsoluteDeviation | 0.000307 | gldm_LargeDependenceEmphasis | 0.000376 |
| firstorder_RobustMeanAbsoluteDeviation | 0.000200 | glrlm_GrayLevelVariance | 0.000537 |
| firstorder_InterquartileRange | 0.000179 | gldm_DependenceVariance | 0.001921 |
| glcm_Imc2 | 0.000159 | glszm_ZonePercentage | 0.000368 |
| glcm_MCC | 0.000130 | ngtdm_Contrast | 0.000598 |
| firstorder_Maximum | 0.000083 | gldm_HighGrayLevelEmphasis | 0.002961 |
| glcm_Idm | 0.000068 | glszm_SmallAreaEmphasis | 0.000108 |
| glcm_Id | 0.000067 | glszm_SizeZoneNonUniformityNormalized | 0.000414 |
| glcm_ClusterShade | 0.000066 | glrlm_ShortRunEmphasis | 0.000124 |
| glcm_Contrast | 0.000065 | glszm_GrayLevelNonUniformityNormalized | 0.003870 |
| glcm_InverseVariance | 0.000060 | ngtdm_Busyness | 0.000561 |
| glcm_DifferenceAverage | 0.000059 | glszm_GrayLevelVariance | 0.000939 |
| firstorder_90Percentile | 0.000013 | glcm_SumSquares | 0.000606 |
| firstorder_Median | 0.000011 | gldm_SmallDependenceEmphasis | 0.000083 |
| firstorder_RootMeanSquared | 0.000010 | glrlm_HighGrayLevelRunEmphasis | 0.003059 |
| firstorder_Mean | 0.000010 | gldm_LargeDependenceLowGrayLevelEmphasis | 0.000149 |

**Table 2. Statistically significant Local Radiomics Features across MGMT 1 and MGMT 0 for (p < 0.05)**

## 6. Bayesian Classification of Activation Maps

The methods outlined in Section 4 of this paper for Bayesian classification of Z-normalized global radiomic features still apply for the classification of local radiomic data with one exception: the input value in this local radiomic use-case is *not* a Z-normalized global radiomic feature value, but rather a feature-specific percentage activation value $\%A$ that is determined on a per-tumor and, thus, per-patient basis. Figure 8 provides bandwidth optimization plots, depictions of $P(x|A) * P(A)$ and $P(x|B) * P(B)$ from equation 7, and the Bayesian classifier's prediction of MGMT 1 or MGMT 0 for a range of hypothetical input $\%A$ values. The examples shown in Figure 8 are the local radiomic analogs to those shown in Figure 6, and they are representative of two Bayesian classifiers developed for two example feature types. Again, the shown plots are for classifiers modelled on the entire dataset of 577 patients, but the 10-fold stratified cross-validation process would utilize 90% of the dataset for modelling during each fold.

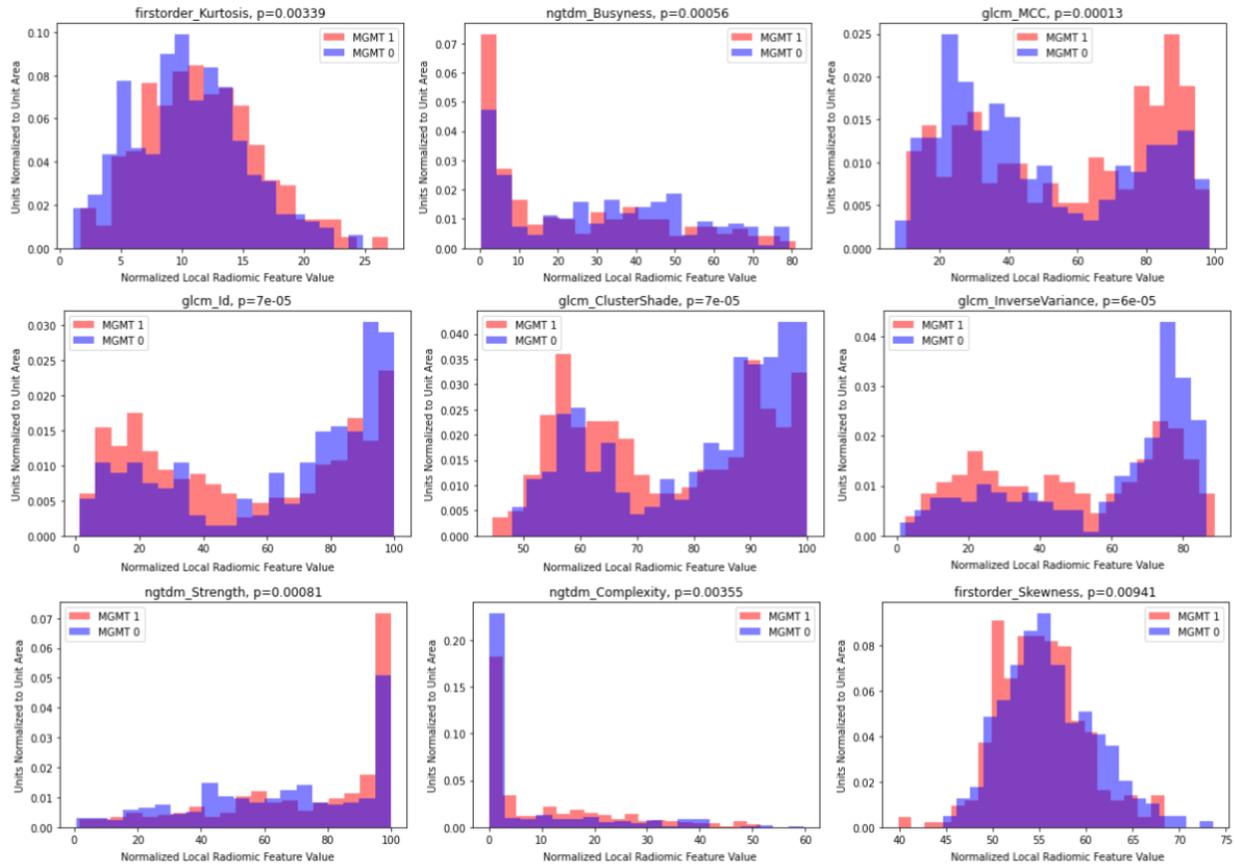

**Figure 7. Statistical Distribution for MGMT 1 and MGMT 0 pairing for Sample Local Radiomics Feature**

## 7. Performance Comparison

In this section, we provide the 10-fold stratified cross-validation results of the global and local radiomics-based Bayesian classifiers for the selected feature types in each case. We then identify the features that were deemed statistically significant in both global and local radiomic approaches, and observe if a local radiomic approach poses any performance benefits in the form of an increase in Bayesian classifier accuracy for those common features.

Figure 9 enumerates the cross-validation results for the global radiomics-based Bayesian classification workflow described in Sections 3 and 4 of this paper. Figure 10 contains the cross-validation results for the novel local radiomic percent activation-based Bayesian classification approach described in Sections 5 and 6 of this paper.

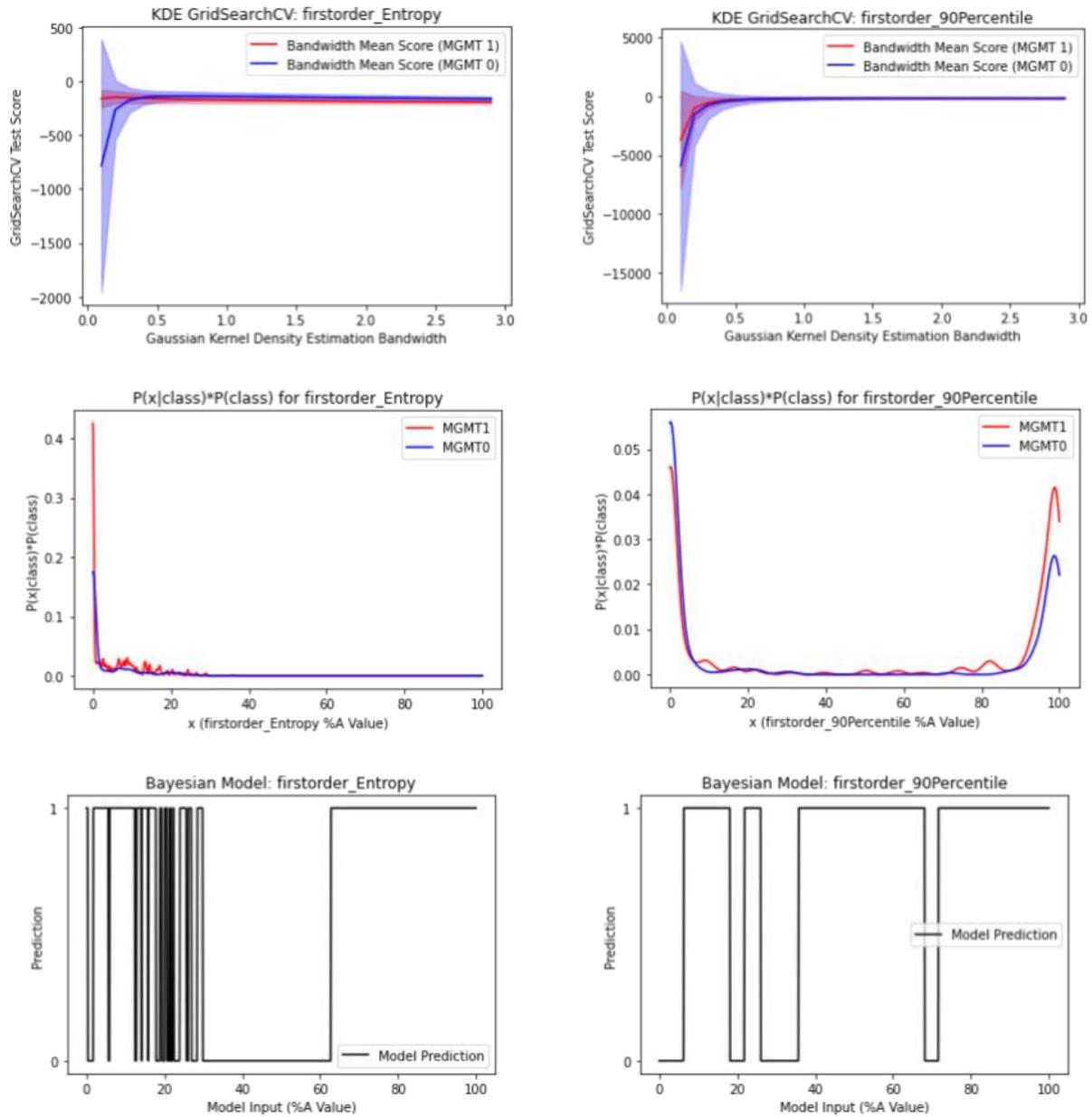

**Figure 8. Bandwidth Optimization and Bayesian Classifier for Local Radiomics Based Predictions**

Figure 11 shows the features that were found to be statistically significant across both the global and local radiomic approaches and provides the percentage change in mean accuracy from global to local radiomics features.

Taking Figures 9 - 11 together, we see that the usage of local radiomic percent activation values as inputs to Bayesian classifiers as opposed to the Z-normalized global inputs consistently yields an increase in mean classifier accuracy across all feature types.

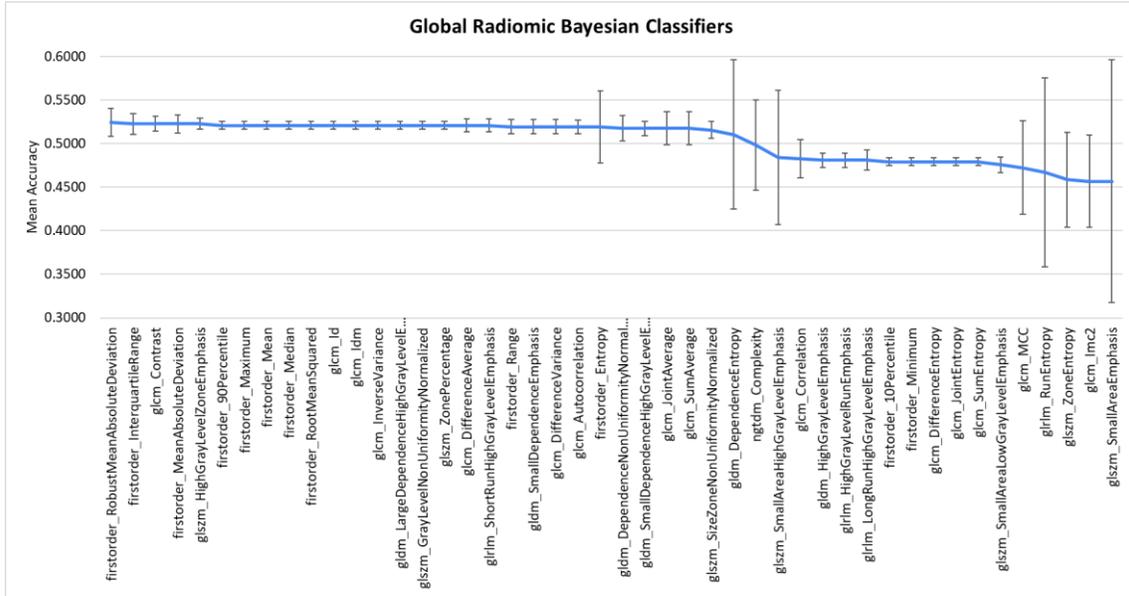

**Figure 9. Mean Cross Validation Accuracy for Global Radiomic Bayesian Classifiers**

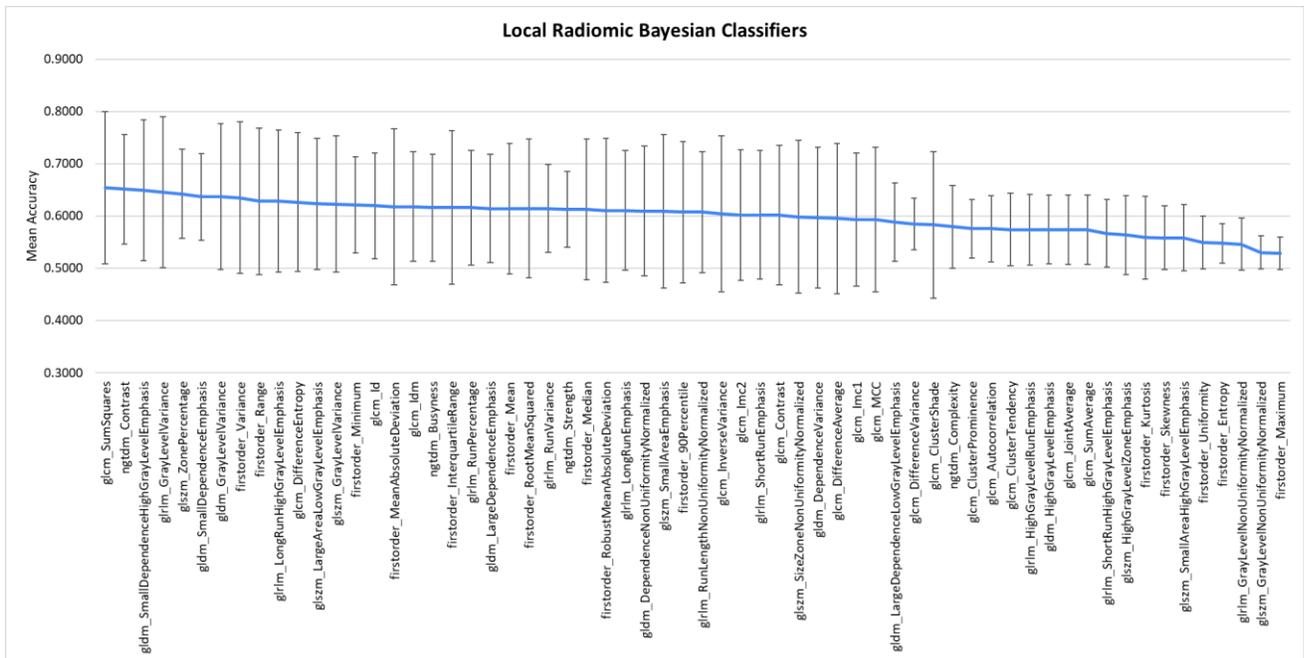

**Figure 10. Mean Cross Validation Accuracy for Local Radiomic Bayesian Classifiers**

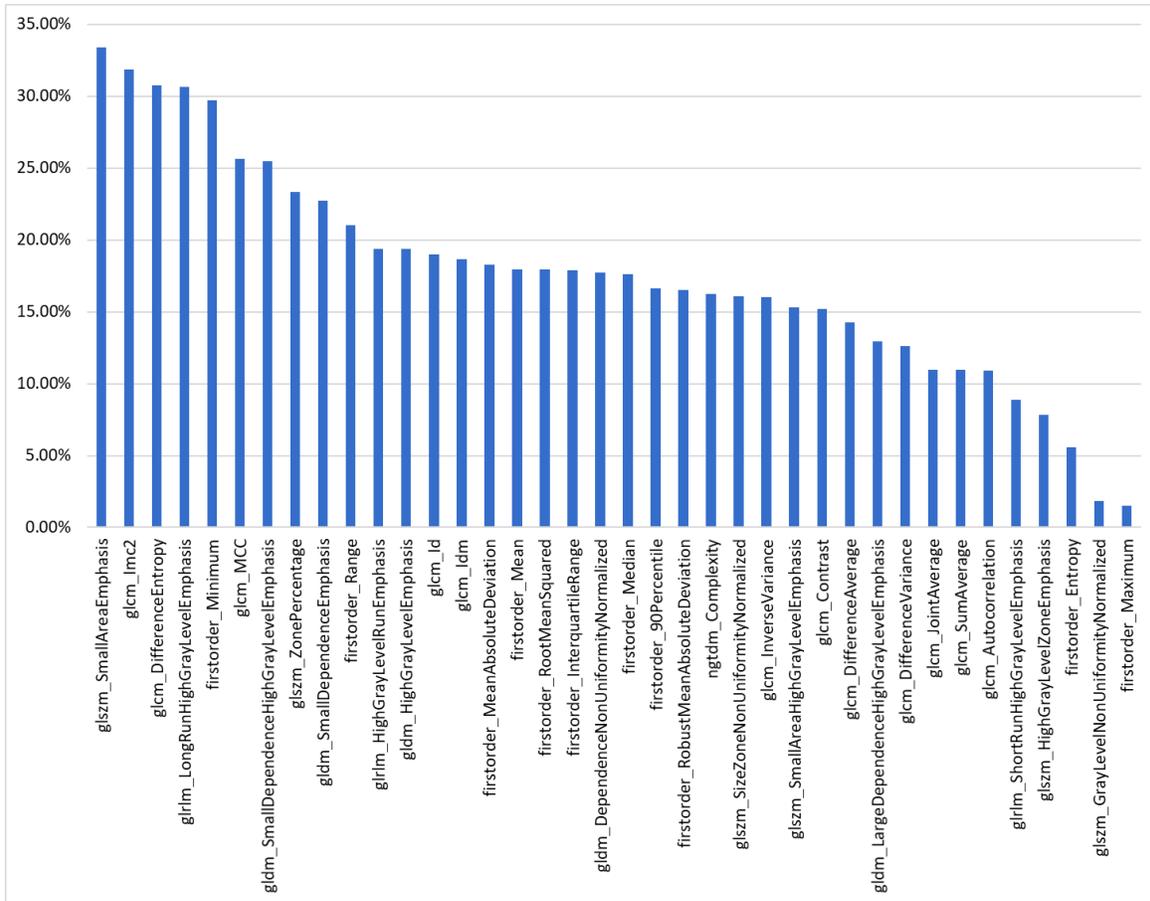

Figure 11. Percentage Increase in Mean Accuracy from Global to Local Radiomic Classifiers

## 8. Conclusion

In this work, we provide several novel insights and methods that advance our understanding of non-invasive MRI-based determination of MGMT methylation status in glioblastoma tumors. To start, we provide the first results identifying radiomic features with statistically significant differences in their distributions for patients with methylated and unmethylated MGMT promoter regions, respectively. Specifically, we identify 46 global radiomic features and 62 local features. Next, we provide the first application of simple Bayesian classifiers to the task of classification based on radiomics data. We demonstrate the boost that local radiomic-based Bayesian classifiers have in comparison to their counterparts that model global radiomic data, and justify further exploration of local radiomic techniques to gain more insightful and accurate MGMT methylation status prediction models. Taken together, this work yields both pure statistical and modelling results that provide a non-invasive means to determine MGMT methylation status in glioblastoma patient's tumors through the comprehensive three-dimensional textural analysis of raw voxel data in MRI scans.